\documentclass{elsart}
\input epsf
\usepackage[dvips]{graphicx}
\usepackage{epsfig}
\usepackage{graphicx}
\usepackage{multicol}
\usepackage{rotating}
\begin{document}

%
%
\begin{frontmatter}
\title{\bf\Large Results and Prospects of \\ Few-Body System Structure 
Studies}
\author {\sf\large Eric~Voutier}
\address{Laboratoire de Physique Subatomique et de Cosmologie \\
IN2P3-CNRS/Universit\'e Joseph Fourier \\ 53 avenue des Martyrs \\ 38026
Grenoble cedex, France}

\vspace*{5pt}

\address{Thomas Jefferson National Accelerator Facility, \\
12000 Jefferson avenue \\ Newport News, Virginia 23606, USA }

%
%
\begin{abstract}
{\small
This paper presents a selected review of the $(e,e'p)$ experimental program 
recently completed at the Jefferson Laboratory in the domain of high momentum
transfer and high recoil momentum. Particular emphasis is put on the current  
understanding of the reaction mechanisms and of the large final state interaction 
effects. Their consequences on the study of the nuclear wave function and 
nucleon-nucleon correlations are addressed. The short and long term future 
of the $(e,e'p)$ experimental program is also discussed in the perspective of 
the study of the quark and gluon structure of nucleons and nuclei. 
}
\end{abstract}
\end{frontmatter}
%
%
\section{Introduction}

Since the very first experiments at Frascati, Saclay, and NIKHEF the $(e,e'p)$ 
reaction has been proven a powerful tool for the investigation of nuclear 
structure~\cite{{Fru84},{Kel96}}. The advent of high energy electron beams with high intensity and 
duty cycle has opened access to the high momentum range, a region expected to 
reveal peculiar features of the nucleon substructure. Indeed, large virtual 
photon momentum $q$ allows access to distances scales $\sim \hbar/q$ that are 
comparable or smaller than the nucleon radius. In addition, the $(e,e'p)$ 
Meson Exchange Current (MEC) mechanism at large $q$ is expected to be less
important due to the natural decrease built into the meson propagators and 
form factors. Therefore, one can expect to probe more reliably high initial 
momenta in the nucleus, that is small inter-nucleon distances, and learn about 
the origin of the short-range repulsion of the NN interaction from its 
standard representation up to the most exotic descriptions. 

The general framework of nuclear structure studies has been evolving since the 
first generation of $(e,e'p)$ experiments at the Continuous Electron Beam 
Accelerator Facility (CEBAF) of the Jefferson Laboratory (JLab). This paper 
reviews our present understanding of the dynamics of the $(e,e'p)$ reaction. 
Particularly, the dominance of the Final State Interactions (FSI) and its 
consequences for the study of the nucleon and nuclear structure are discussed. 
Future prospects in this domain as well as more exotic use of the $(e,e'p)$ 
reaction in the study of the Color Transparency (CT) phenomenon are also 
addressed. 

The next section introduces the basic concept and description of the $(e,e'p)$ 
reaction. Recent results on D~\cite{Ulm02}, $^3$He~\cite{Rva04}, and $^4
$He~\cite{Rei03} are then discussed which allow one to envisage more sensitive 
investigations of the CT phenomenom~\cite{JLa12}. The problem of the left-right 
asymmetry observable is then discussed on the basis of a comparison between 
few- and many-body experimental results~\cite{{Rva04},{Gao00}}. In 
sec.~\ref{sec:NNcor}, the most recent investigations of nucleon-nucleon 
correlations~\cite{{Niy04},{Ben04}} in the nuclear medium are presented. Finally, 
the long standing issue of bound nucleon form facors is revisited in light of 
the new JLab measurements~\cite{Str03}.

%
%
\section{The ${\mathbf (e,e'p)}$ Reaction}
\label{sec:eep}

\begin{figure}[h]
\begin{center}
\epsfig{file=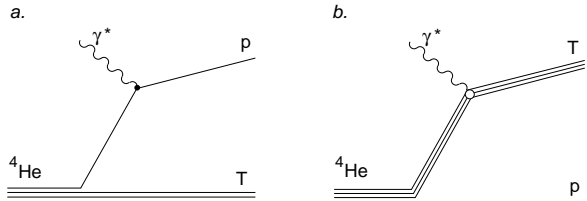, width=220pt}
\caption{One-body contributions to the $^4$He$(e,e'p)$ reaction: plane wave
impulse approximation (a), and recoil contribution (b).}
\label{PWIA}
\end{center}
\end{figure}

\begin{figure}[h]
\leftline{\hspace*{10pt} \epsfig{file=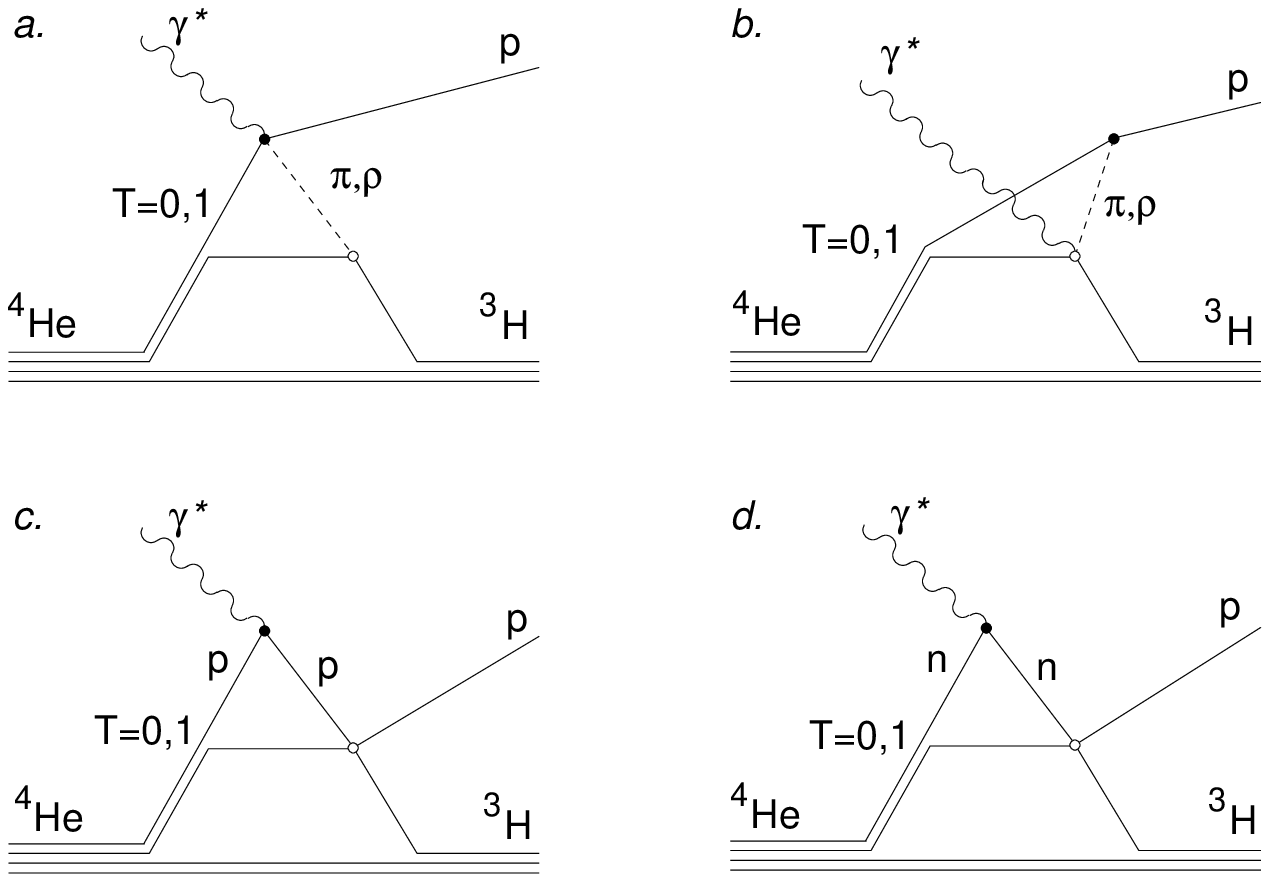, width=188pt}}

\vspace*{-133.5pt}

\rightline{\epsfig{file=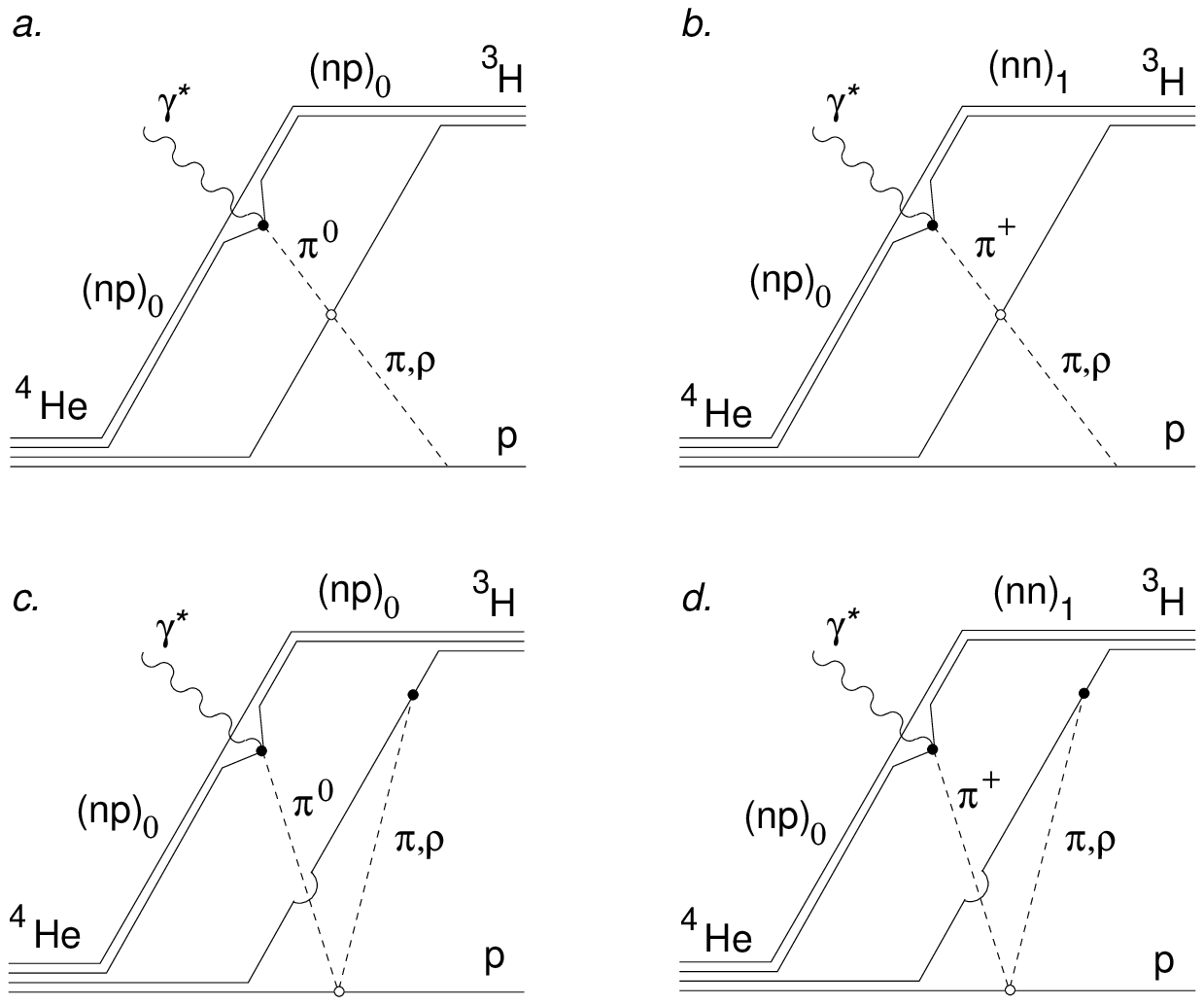, width=160pt} \hspace*{10pt}}
\caption{Two-body (four left diagrams) and three-body (four right diagrams) 
contributions to the $^4$He$(e,e'p)$ reaction: final state interaction (c and d 
of left panel), and meson exchange currents (all other diagrams).}

\label{2-3body}
\end{figure}

\vspace*{1pt}
The merit of quasi-elastic electron scattering for the study of nuclear
structure is in the description of the reaction process in the framework of the 
Plane Wave Impulse Approximation (PWIA). In this context (fig.~\ref{PWIA}a), the 
cross section for the $(e,e'p)$ process can be factorized in two terms: one 
representing the probability of a virtual photon to interact with a proton, that 
is the electron-bound proton cross section $\sigma_{ep}$~\cite{For83}, and 
another representing the probability to find a proton in the nucleus with initial 
momemtum ${\vec p}_m$ and binding energy $E_m$, that is the nuclear spectral 
function 
$S(p_m,E_m)$~\cite{{Fru84},{Kel96}}. Tagging the virtual photon $({\vec q},\omega) 
= ({\vec k}-{\vec k}',E-E')$ by measuring the four-momenta $({\vec k},E)$ and 
$({\vec k}',E')$ of the incoming and scattered electron, and measuring the 
four-momentum $({\vec p},E_p)$ of the knocked-out proton, one reconstructs
\begin{eqnarray}
{\vec p}_{r} & = & {\vec q} - {\vec p} \label{prequa}\\
E_m & = & \omega - T_p - T_r \ . 
\end{eqnarray} 
Within PWIA, the momentum ${\vec p}_{r}$ of the recoil system is equal to the
opposite of the initial momentum ${\vec p}_m$ of the proton. Then, studying the dependence of the 
cross section either on $(q,\omega)$ or on $(p_m,E_m)$, one can access directly 
different features of the nuclear structure. Particularly, at high initial 
momentum, one can determine the importance and the origin of high momentum 
components of the nuclear wave function, test the existence of relativistic 
effects or learn about short range correlations. At high momentum transfer, one 
can question the validity of the current operator or investigate the 
electromagnetic properties of bound nucleons, and possible exotic configurations. 
\newline
However, depending on the kinematics of the reaction, several other mechanisms 
interfere with the elementary PWIA amplitude. FSI, where the knocked-out proton 
interacts with the residual nucleus (fig.~\ref{2-3body}) distort the nuclear 
information and break the simple factorization scheme of the cross 
section. In addition, instead of interacting with nucleons, the photon can 
couple to the virtual mesons of the nuclear field leading to a large variety of 
MEC contributions (fig.~\ref{2-3body}). 

\vspace*{1pt}

One way to disentangle the different reaction mechanisms 
and select the PWIA amplitude is to separate the elementary components or 
response
functions of the cross section and to take advantage of peculiar kinematics
conditions. For instance, MEC essentially affect the transverse components of 
the hadronic current and their effects are expected to decrease at high 
momentum transfers because of the natural $Q^{-2}$ evolution of meson form 
factors. 

The unpolarized cross section for the A$(e,e'p)$ process may be written
\begin{equation}
\hspace{-15pt}\frac{{\mathrm d}^6\sigma}{{\mathrm d}{\vec k}'{\mathrm d}
{\vec p}} = \vert {\vec p} \,\vert E_p \, \sigma_M\,\bigg[ V_L R_L + V_T R_T + 
V_{LT} R_{LT} \cos{(\phi)} + V_{TT} R_{TT} \cos{(2\phi)} \bigg]
\end{equation}
\begin{figure}[ht]
\begin{center}
\epsfig{file=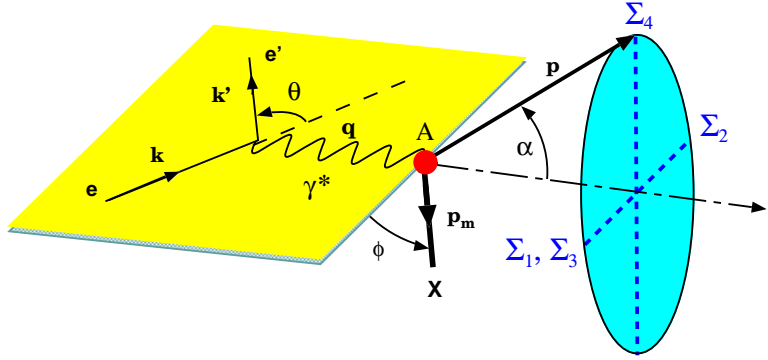, width=290pt}
\caption{The leptonic and hadronic planes for the $(e,e'p)$ process.} 
\label{reaction}
\end{center}
\end{figure}
where $\sigma_M$ is the Mott cross section for a structureless nucleus
\begin{equation}
\sigma_M = \alpha^2  \, {\left( \frac{1}{2 E \sin(\theta/2)} \right)}^2 
\, \tan^2 \left( \frac{\theta}{2} \right) 
\end{equation}
with $\theta$ the electron scattering angle and $\phi$ the angle between the 
leptonic and hadronic planes (fig.~\ref{reaction}). The $V_i$ are the 
kinematical factors 
\begin{eqnarray}
V_L & = & \frac{Q^4}{q^4} \\
V_T & = & \frac{1}{2} \frac{Q^2}{q^2} + \tan^2 \left(\frac{\theta}{2}\right) \\
V_{TT} & = & \frac{1}{2} \frac{Q^2}{q^2} \\
V_{LT} & = & \frac{Q^2}{q^2} \left[ \frac{Q^2}{q^2}+ \tan^2 \left(\frac{\theta}{2} 
\right) \right]^{1/2}
\end{eqnarray}
depending on the virtual photon four-momentum characteristics; $Q^2$ the 
relativistic invariant $q^2-\omega^2$, by convention. The $R_i$ represent the 
response of the nuclear system to the electromagnetic excitation and can be
expressed as   
\begin{eqnarray}
R_L & = & \rho \rho^{\dagger} \\
R_T & = & J_xJ_x^{\dagger}+J_yJ_y^{\dagger} \\
R_{TT} \, \cos(2\phi) & = & J_xJ_x^{\dagger}-J_yJ_y^{\dagger} \\
R_{LT} \, \cos(\phi) & = & -(\rho J_x^{\dagger}+J_x\rho^{\dagger})
\end{eqnarray}
in terms of the hadronic current $({\vec J},\rho)$ in the reference frame of 
the virtual photon (fig.~\ref{reaction}). The separation of these four response
functions requires four different measurements ($\Sigma_i$ on
fig.~\ref{reaction}): measuring the cross section on 
the right ($\Sigma_1$) and  left ($\Sigma_2$) of the virtual photon at constant 
$({\vec q},\omega)$ and $({\vec p},E_p)$, one extracts the 
longitudinal-transverse interference response $R_{LT}$; an additionnal 
measurement out of the reaction plane ($\Sigma_4$) allows one to obtain the
transverse-transverse interference response $R_{TT}$; a last measurement at a 
different beam energy ($\Sigma_3$) allows one to separate the longitudinal 
$R_L$ and transverse $R_T$ responses by changing the electron scattering angle.  

Following this philosophy, several $(e,e'p)$ experiments on few-body systems 
at high momemtum transfer and high recoil momentum have been completed using
the High Resolution Spectrometers~\cite{Alc04} of the hall A of JLab. Few-body 
systems are indeed priviledged laboratories since exact microscopic calculations 
can be achieved involving the full complexity of the NN interaction.

%
%
\section{High Momenta Cross Sections}
\label{sec:Xsection}

Beyond the experimental benefit of the large duty factor of CEBAF, the several 
GeV beam energy allows one to access very small cross sections typical of the 
high recoil momentum region. Therefore, the cross section can be mapped over a 
large range of recoil momentum without changing the virtual photon probe. This 
feature is an important improvment with respect to older experiments at smaller 
scale facilities: keeping the leptonic vertex onstant gives a better handling 
of the reaction mechanisms and consequently insures a better interpretation of 
experimental data. 

\vspace*{-10pt}
\begin{figure}[hbt]
\leftline{\hspace*{-10pt}\epsfig{file=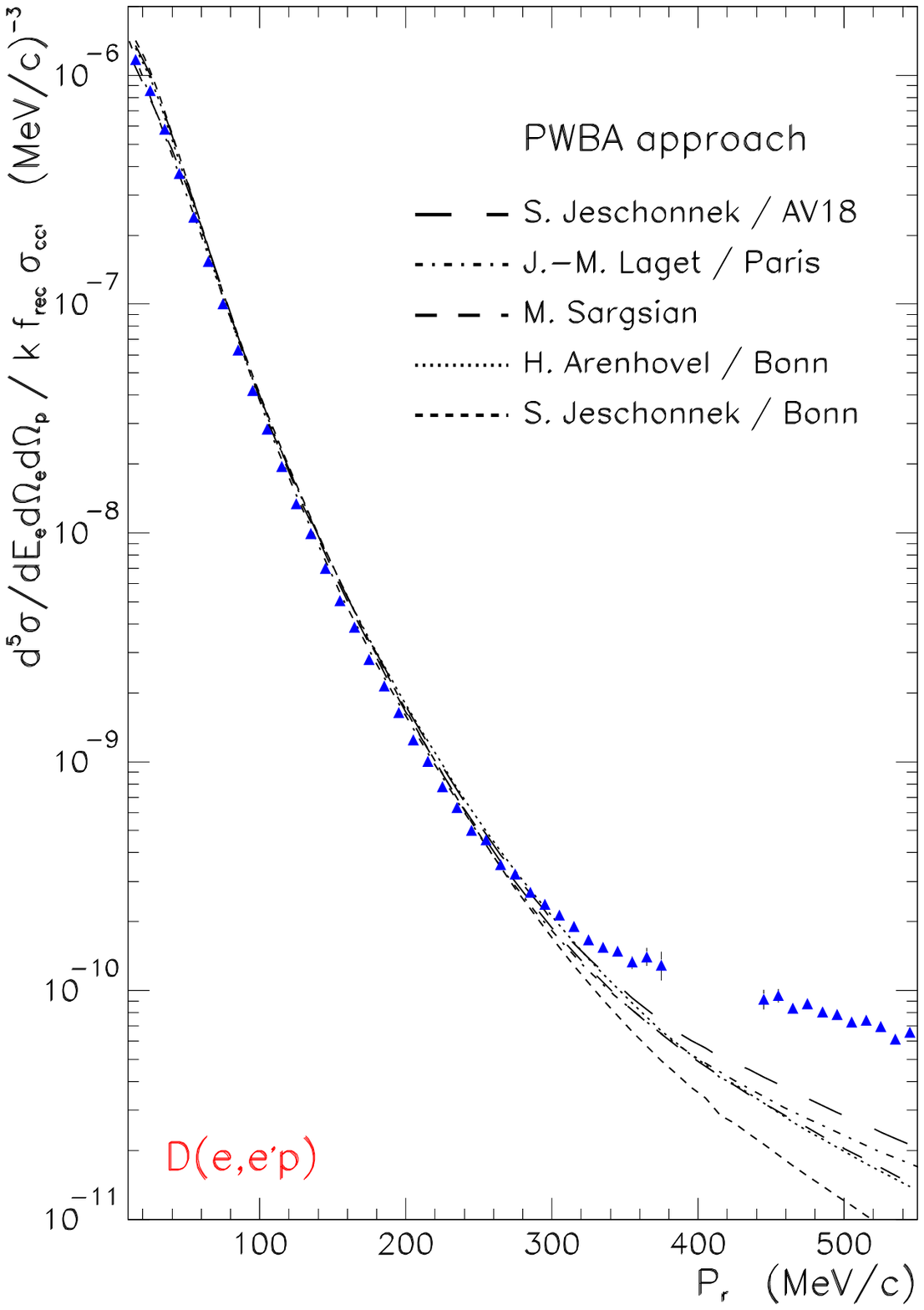, width=240pt}}

\vspace*{-217pt}

\rightline{\epsfig{file=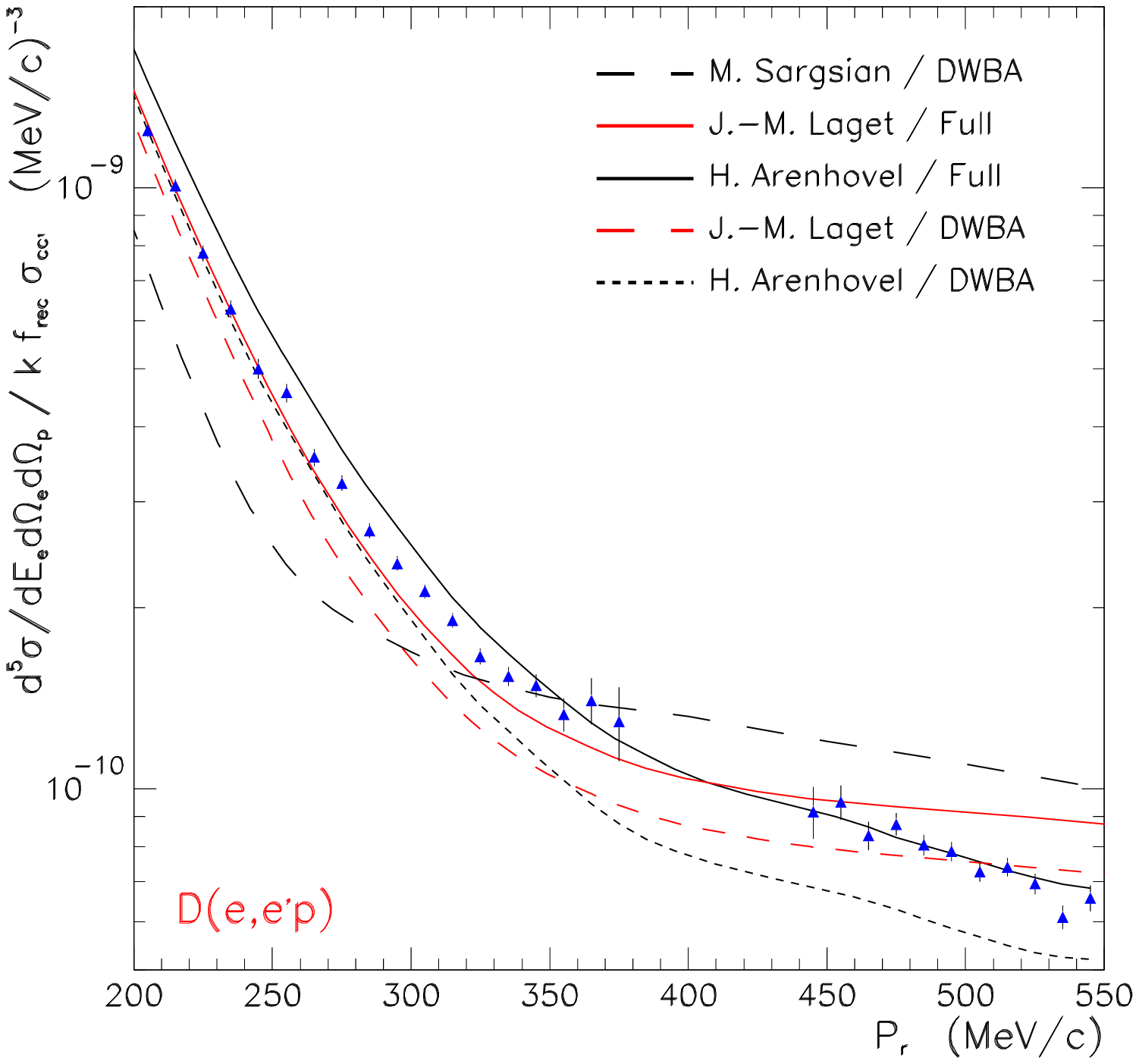, width=240pt}\hspace*{-10pt}}
\vspace*{-33pt}
\caption{Experimental data of the E94-004 experiment~\cite{Ulm02}: comparison 
to PWIA calculations (left) over the full range of measured $p_r$, and 
comparison to full calculations (right) in the high momentum region}
\label{ulmer}
\end{figure}

\subsection{D$(e,e'p)$ cross section}
\label{sec:deuton}

The first experiment reported here is about deuterium, the elementary 
laboratory nucleus of the NN interaction. This D$(e,e'p)$ electro-disintegration 
experiment~\cite{Jon94} has been performed in quasi-elastic kinematics at 
$Q^2=0.67$~GeV$^2$ and $x = Q^2 / 2 M_p \omega = 0.96$. The recoil neutron is
reconstructed (eq.~\ref{prequa}) perpendicular to the virtual photon. 
Experimental data are compared in fig.~\ref{ulmer} to several 
calculations. It is particularly shown that all PWIA calculations 
using different methods and/or nuclear wave functions fail to reproduce the data 
at large recoil momentum. Above 300~MeV/$c$, FSI drive the cross section and 
small but still active MEC have to be taken into account as shown by 
calculations from H.~Arenh\"ovel~\cite{Ulm02} and J.-M.~Laget~\cite{Lag04}.

\subsection{$^3$He$(e,e'p)$ cross section}
\label{sec:2bbu}

\begin{figure}[htb]
\leftline{\epsfig{file=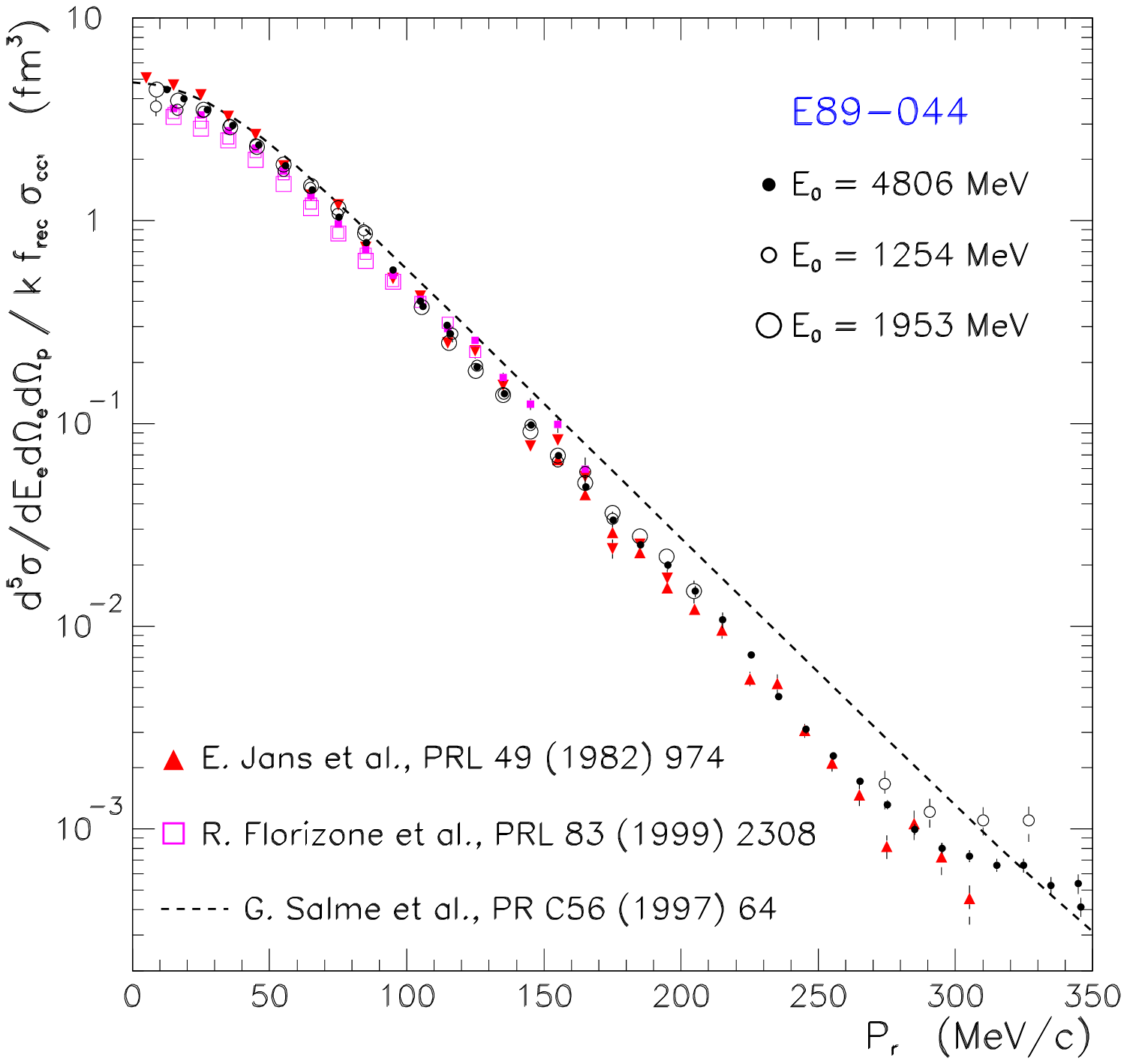, width=220pt}}

\vspace*{-220pt}

\rightline{\epsfig{file=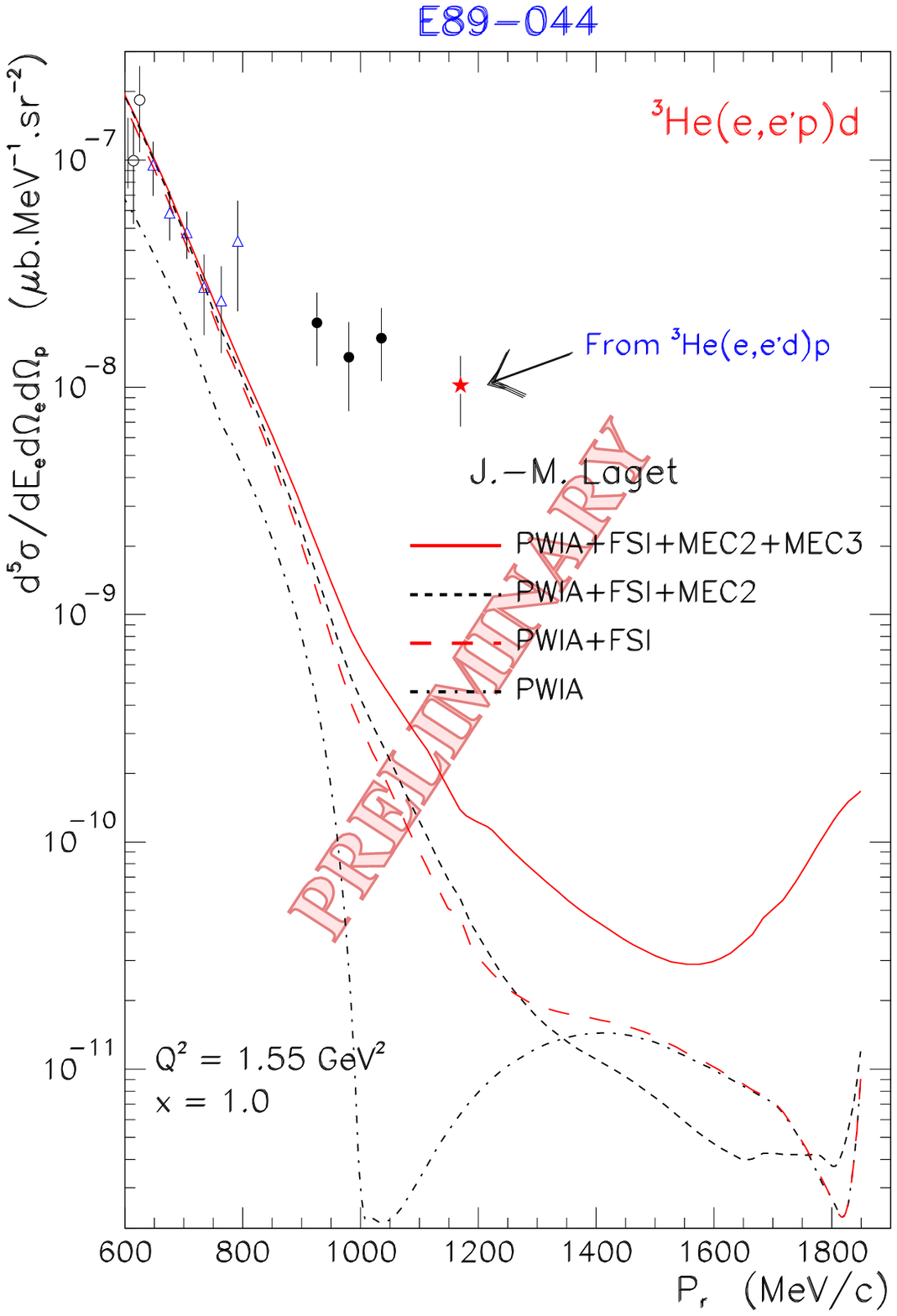, width=220pt}}

\vspace*{-60pt}

\begin{center}
\epsfig{file=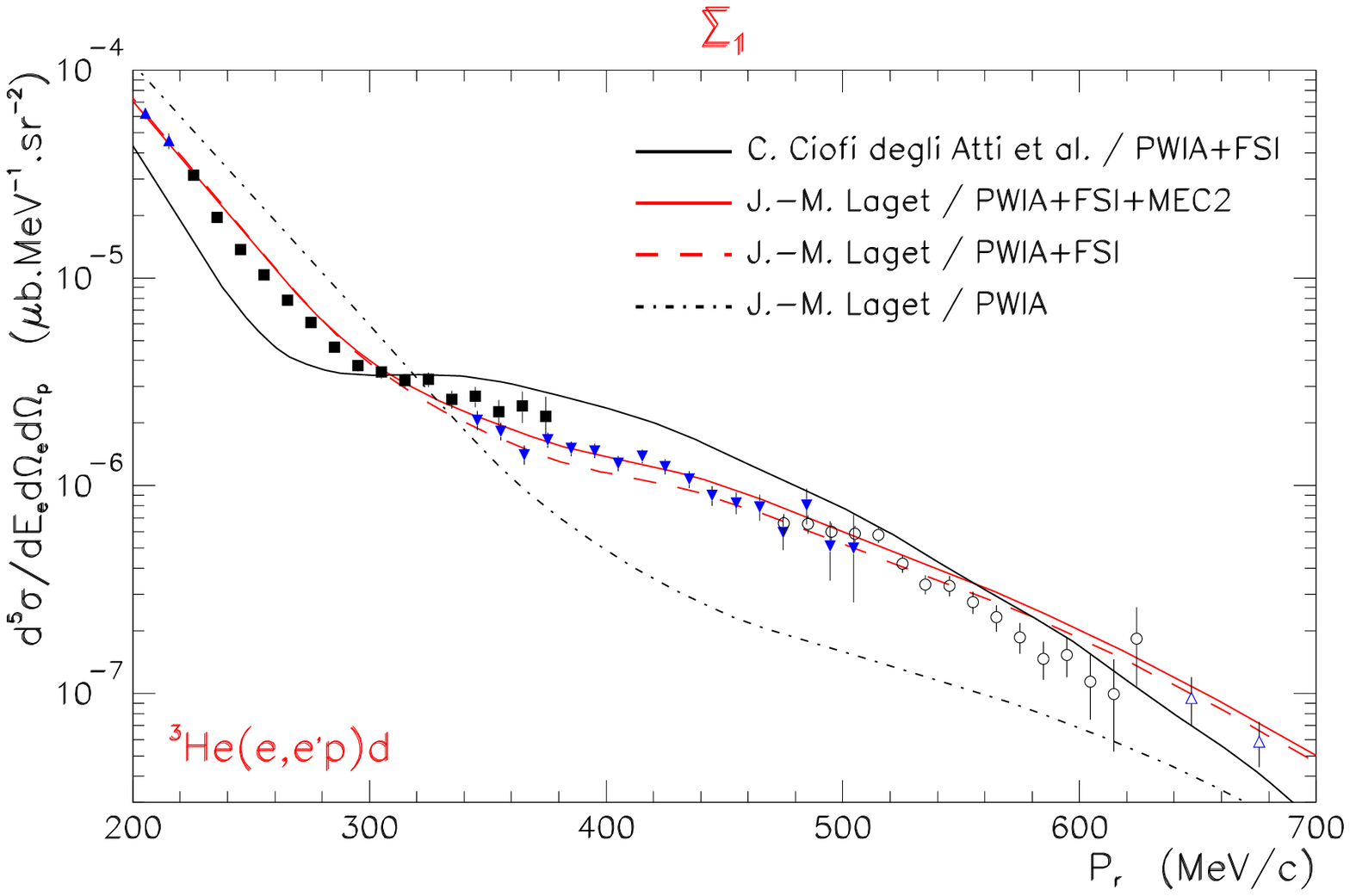, width=280pt}
\vspace*{-60pt}
\caption{Selected data of the E89-044 experiment~\cite{Rva04}: comparison to
previous experiments (top left), comparison to model calculations in the high
(bottom) and very high (top right) recoil momentum region.}
\label{marat}
\end{center}
\end{figure}

Because of its higher density, the helium nucleus is expected to reveal more 
fine details of the NN interaction. It is also a bridge between few- and 
many-body systems where one can investigate the strength of three-body forces. 

A selected set of data from the $^3$He$(e,e'p)d$ JLab experiment~\cite{E89044} 
is shown in fig.~\ref{marat}. The experiment was performed in perpendicular 
kinematics at constant $Q^2=1.52$~GeV$^2$ and $x=0.99$, reaching for the very 
first time the 1~GeV/$c$ recoil momentum range. It corresponds to an 
unprecedented  measurement of the cross section over six orders of magnitude.
\newline
The top left panel of fig.~\ref{marat} represents the reduced cross section for the 
three beam energies of the experiment as compared to older 
results~\cite{{Jan82},{Flo99}} and a PWIA calculation~\cite{Kie97}; within PWIA, 
the reduced cross section corresponds to the nuclear spectral function 
$S(| {\vec p}_m | = |{\vec p}_r|,E_m=5.4\,{\mathrm {MeV}})$. Good agreement with 
older data is obtained 
up to 250~MeV/$c$. Above this value, data dispersion is observed corresponding, 
for different beam energies, to different magnitude of the reaction mechanisms 
beyond PWIA. This is confirmed by the comparison to a PWIA calculation where the 
agreement degrades gradually starting about 100~MeV/$c$. \newline
Similarly to the deuterium case, the high momentum range (fig.~\ref{marat} bottom 
panel) exhibits strong effects from the interaction of the knocked-out proton 
with the residual system: below 300~MeV/$c$, the PWIA cross section is moderately 
quenched while above this value strong enhancement of the cross section (about 
a factor 5 at 500~MeV/$c$) is observed. Different calculations~\cite{{Lag03},{Cio04}} 
confirm that FSI is the driving contribution, MEC being quasi-negligible, as 
expected in this $Q^2$ range. Note that, presented at this workhsop, new 
calculations~\cite{Udi04} within a Relativistic Distorted Wave Impulse 
Approximation (RDWIA) show good agreement with data and lead to the same 
conclusion about the dominance of FSI in this energy range.\newline 
In the very high momentum region (fig.~\ref{marat} top right), which was explored 
for the very first time, the cross section seems to saturate at an unexpectedly 
high level. These data are confirmed by the measurement of the recoil deuteron in 
the same experiment~\cite{Maz03}: within a simple jacobian transformation, the 
$^3$He$(e,e'p)d$ cross section is deduced from the measured $^3$He$(e,e'd)p$ 
cross section. Calculations are unable to reproduce this excess of strength at 
1~GeV/$c$. Whether it is a consequence of the truncation of the diagrammatic 
expansion~\cite{Lag04} or a signature of other degrees of freedom is an open 
question. \newline
More data on this experiment are available in~\cite{Rva03}, particularly a 
separation of the response functions was achieved which however does not bring 
additional information on the nuclear structure due to the magnitude of FSI 
effects.

\subsection{$^4$He$(e,e'p)$ cross section}

\begin{figure}[h]
\begin{center}
\epsfig{file=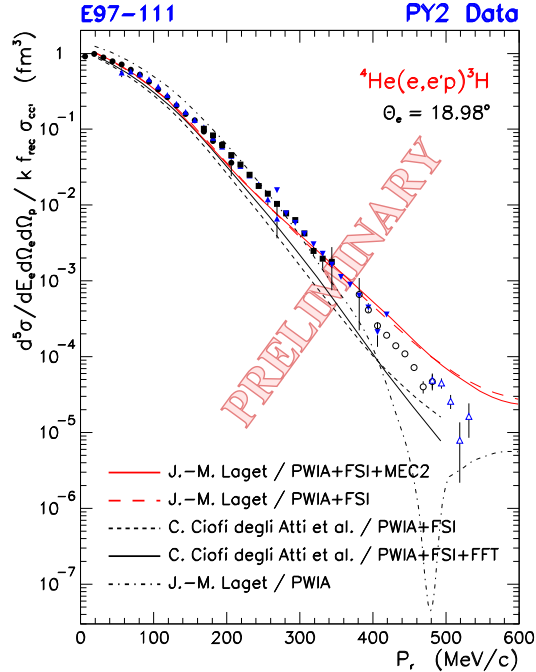, width=200pt}
\caption{Selected preliminary data of the E97-111 experiment~\cite{Rei03} 
measured for a 3.170~GeV beam energy and a fixed electron scattering angle 
$\theta_e = 18.98^{\circ}$: the different curves are several model 
calculations.}
\label{bodo}
\end{center}
\end{figure}

Another attempt to pin down the nuclear wave function at high recoil momentum 
was investigated in a $^4$He$(e,e'p)^3$H experiment~\cite{E97111}. It was shown 
by S.~Tadokoro et al.~\cite{Tad87} that this nucleus is particularly sensitive 
to the details of the short range part of the NN interaction: within an 
independent particle approach, computing the spectral function from a realistic 
potential, they found that it shows a typical diffractive pattern which minima 
position and 
height are sensitive to the fine structure of the NN interaction. The expected 
position at about 450~MeV/$c$ lies in the high momentum range and it is not 
surprising that a previous experiment at smaller beam energy~\cite{Lee98} was 
unable to observe it, suffering from large FSI and MEC effects that fill the 
dip. 

The E97-111 experiment took data in parallel kinematics which is expected 
to minimize FSI. Here, the knocked-out proton and the recoil system are emitted 
along the direction of the virtual photon. Because data were taken at constant 
beam energy and scattered electron angle, the momentum evolution of the cross 
section mixed different $({\vec q},\omega)$, as opposed to the previously
reported experiment. Experimental data are compared to theoretical calculations 
in fig.~\ref{bodo}: they all show large FSI effects which forbid observation of 
the minimum. 

In conclusion, these experiments, which are relatively well understood by
theoretical calculations using modern nuclear wave functions, show the dominance
of FSI at high recoil momentum. These large effects consequently forbid us to 
go beyond the actual knowledge of the NN interaction and learn about the fine 
details of the wave function. There are however theoretical expectations and 
experimental indications~\cite{{Rei03},{Pen04}} that at large $x$, when the 
direction of the initial proton momentum is opposed that of the virtual photon,  
the magnitude of FSI should decrease, enabling the selection of the PWIA
amplitude. 

%
%
\section{Final State Interactions and Color Transparency}

The main process at work in the interaction of the knocked-out proton with the 
residual system is the NN scattering. Because of the large energy transfer in 
JLab experiments, the proton can rescatter on a nucleon at rest which then 
gains momentum and is emitted at about 90$^{\circ}$ with respect to the proton. In this way, part 
of the strength at small initial momentum is shifted towards high recoil 
momentum. FSI are then maximal at large $p_r$ in perpendicular kinematics. More 
strictly, the position of this maximum corresponds to the singularity of the 
nucleon propagator which occurs at $x=1$ in the $(e,e'p)$ channel~\cite{Lag04}: the 
electron scatters on a proton which propagates on-shell in the nuclear medium 
and rescatters on a nucleon at rest. Since the PWIA amplitude falls rapidly 
with $p_r$, the on-shell rescattering takes over and dominates at large $p_r$. 

Several theoretical treatments of FSI exist in the literature ranging from 
microscopic to macroscopic approaches: in the diagrammatic approach of
J.-M.~Laget, FSI are parametrized in terms on the NN partial wave expansion 
for $\omega < 500$~MeV, and in terms of a high energy parametrization fitted to 
the NN scattering world data at larger excitation energy~\cite{Lag04}; Glauber 
based calculations are also available~\cite{{Cio04},{Bia96}} as well as a
Generalized Eikonal Approximation~\cite{Fra97} that takes into account the 
Fermi motion of the target nucleon, as opposed to Glauber; optical potential 
approaches have also been developped~\cite{Udi04}. 

\begin{figure}[ht]
\leftline{\hspace*{1pt} \epsfig{file=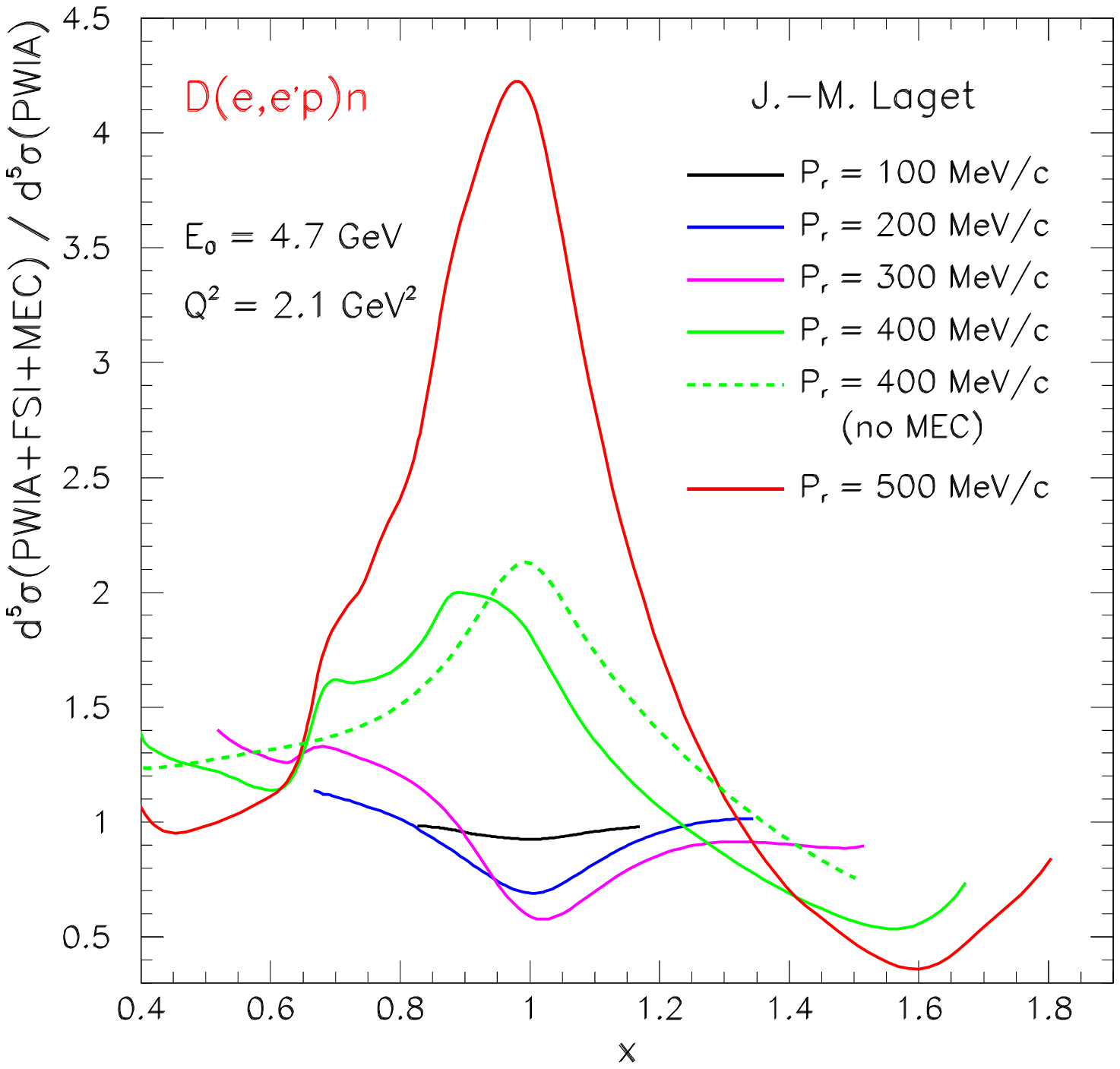, width=190pt}}

\vspace*{-191pt}

\rightline{\epsfig{file=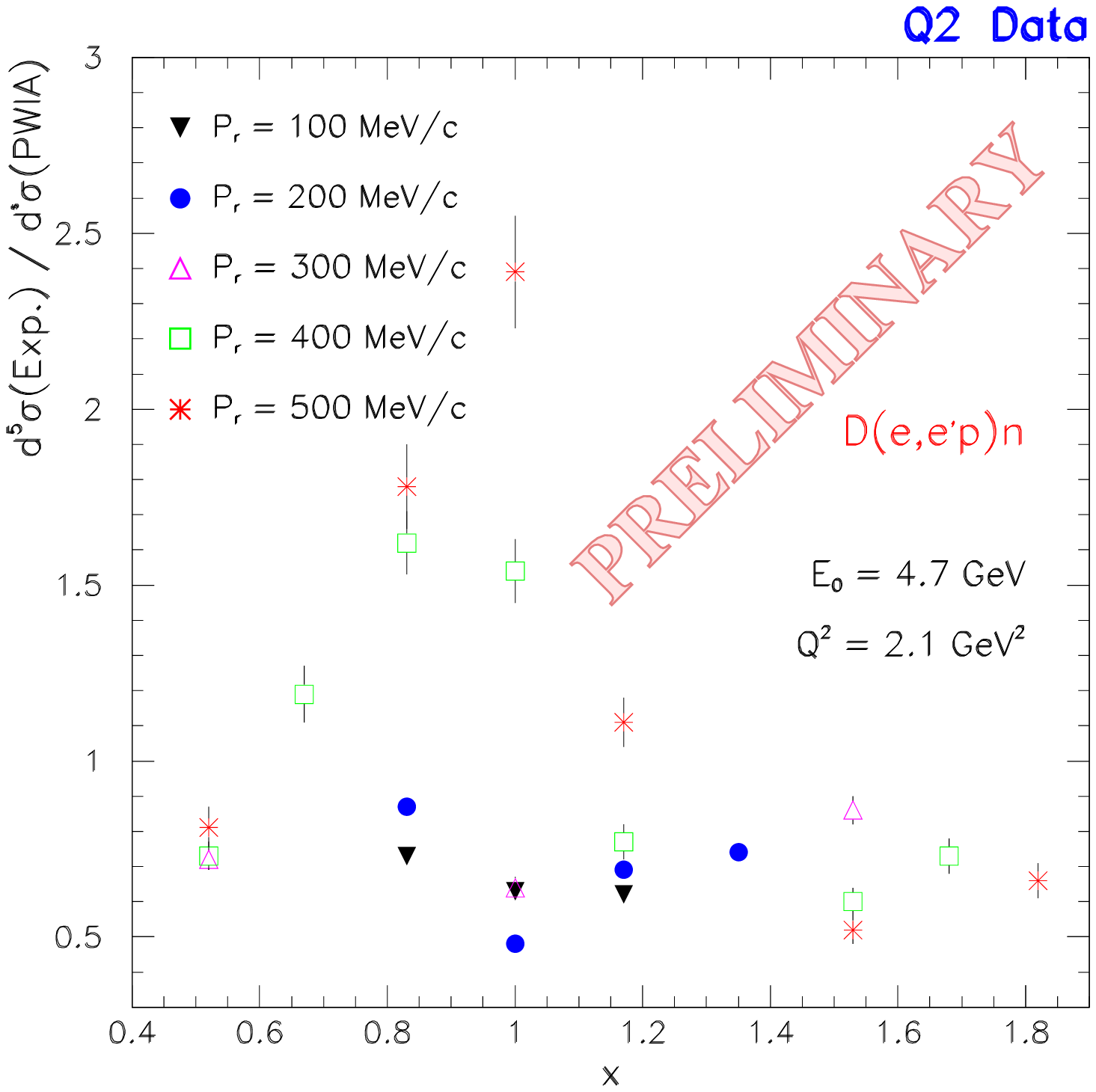, width=190pt} \hspace*{1pt} }

\caption{Selected preliminary data (right panel) of the E01-020 
experiment~\cite{Boe03} at $Q^2=2.1$~GeV$^2$ and several recoil momenta, 
together with corresponding theoretical calculations~\cite{Lag04} (left panel).}
\label{werner}
\end{figure}

This important theoretical activity as well as the still open question of the 
origin of the high momentum components of the wave function has motivated a 
systematic study of the $(e,e'p)$ reaction in the elementary deuterium 
nucleus. The E01-020 experiment~\cite{E01020} has measured the angular 
distribution of the recoil neutron (or $x$ distribution~\footnote{At fixed $Q^2$
and $p_r$, the angle of the recoil particle and the $x$ variable are linked via 
a unique relationship, large $x$ corresponding to small recoil angles.}) in 
the D$(e,e'p)$ reaction over the complete allowed phase space at fixed $Q^2$ 
and $p_r$. Preliminary data~\cite{Boe03} at $Q^2=2.1$~GeV$^2$ and several 
recoil momenta are reported in fig.~\ref{werner} in terms of the ratio of the
measured experimental cross section to a model calculation of the PWIA 
amplitude. A point-to-point comparison to theoretical calculations should not 
be attempted here because of the very preliminary stage of the analysis, only 
the general shape and behaviour of the data is relevant. The general trend of 
data is in agreement with calculations, showing a moderate FSI quenching of the 
cross section at small $p_r$ and an impressive FSI enhancement at large $p_r$. 
At 400~MeV/$c$ recoil momentum, the peak position is shifted below $x=1$ which, 
similarly to the on-shell nucleon mechanism, corresponds to the propagation of 
an on-shell $\Delta$ resonance in the nuclear medium. 

\vspace*{15pt}
This simple on-shell rescattering mechanism which relies on elementary on-shell 
amplitudes opens an original use of the $(e,e'p)$ reaction, namely, the study 
of the CT phenomenom in few-body systems~\cite{{Vou96},{Lag97}}. \newline
Color Transparency is a direct consequence of Quantum Chromo-Dynamics that
predicts the existence, in the nucleon wave function, of a minimal valence 
state where the quarks are very close together and constitute a small size 
color neutral object (or mini-hadron). Such a color singlet system cannot emit 
or absorb soft gluons and therefore experiences much reduced strong interaction 
with other nucleons when travelling through the nuclear medium. 

\begin{figure}[ht]
\begin{center}
\epsfig{file=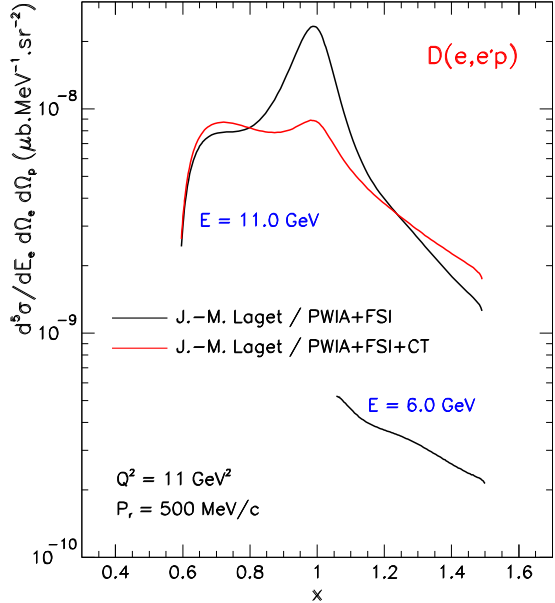, width=210pt}
\vspace*{-10pt}
\caption{Predictions for the $x$ distribution of the D$(e,e'p)$ cross section 
at fixed $Q^2=11$~GeV$^2$ and $p_r=500$~MeV/$c$ for two different beam 
energies: CT effects (red curve) result in a reduction of the cross section at 
$x=1$.}
\label{colo12}
\end{center}
\end{figure}
The experimental evidence of CT not only requires the selection of a small size 
configuration but also a clear signature of the subsequent reduced interaction. 
As of now, the several experimental attempts~\cite{{One95},{Abb98},{Gar02}} to 
search for CT in many-body systems have failed. The common understanding is that 
the coherence length, that is the distance needed for the mini-hadron to evolve 
from its minimal valence state toward its asymptotic wave function, is too 
small with respect to the typical nuclear scale. \newline
Few-body systems may solve this problem by taking advantage of the dominance of FSI 
in perpendicular quasi-elastic kinematics: the signature of CT would appear as a 
decrease of the cross section at $x=1$ as a function of $Q^2$, due to the gradual 
suppression of FSI with diminishing mini-hadron size. A similar example is shown on
fig.~\ref{colo12} in the context of the 12~GeV upgrade project at 
JLab~\cite{JLa12}. The merits of the energy upgrade to access a wide $x$ range 
at large $Q^2$ is obvious. This particularly allows the study of the D$(e,e'p)$ 
reaction in the quasi-elastic region where the existence of CT would lead to 
large effects on the cross section.

%
%
\section{The ${\mathbf A_{LT}}$ Dilemna}
 
As explained in sec.~\ref{sec:eep}, the study of $(e,e'p)$ reactions at JLab is 
not restricted to cross section measurements. Some experiments aimed at and 
achieved a separation of the response functions~\cite{{Gao00},{Rva03}}. 
Particularly, the longitudinal-transverse interference response was extracted. 
From the same data, one can also reconstruct the $A_{LT}$ observable 
(see fig.~\ref{reaction})
\begin{equation}
A_{LT} \, = \, \frac{\Sigma_2-\Sigma_1}{\Sigma_2+\Sigma_1} \, = \, \frac{V_{LT} 
R_{LT}}{V_L R_L + V_T R_T + V_{TT} R_{TT}}
\end{equation}
which corresponds to the left-right asymmetry with respect to the virtual 
photon. This observable downplays the significance of the ground-state wave
function by virtue of the ratio involved in its definition~\cite{Kel96}. There 
exists indications that it is sensitive to relativistic effects at small 
momentum ($p_r<200$~MeV/$c$)~\cite{Gil98}, and more generally to any mechanism 
that breaks the simple factorization scheme of the cross section~\cite{Udi01}. 
  
\begin{figure}[h]
\leftline{\hspace*{3pt} \epsfig{file=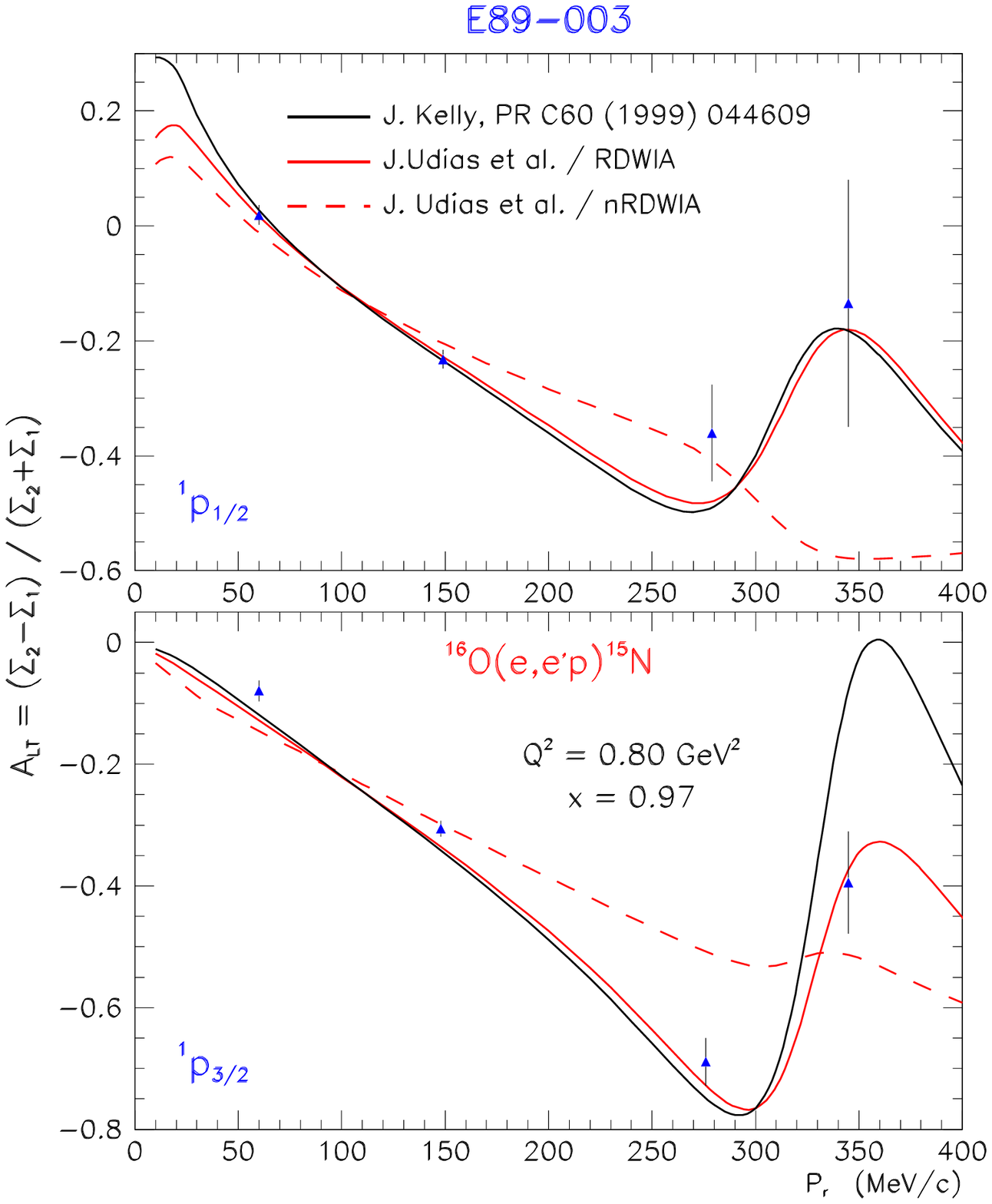, width=190pt}}

\vspace*{-232pt}

\rightline{\epsfig{file=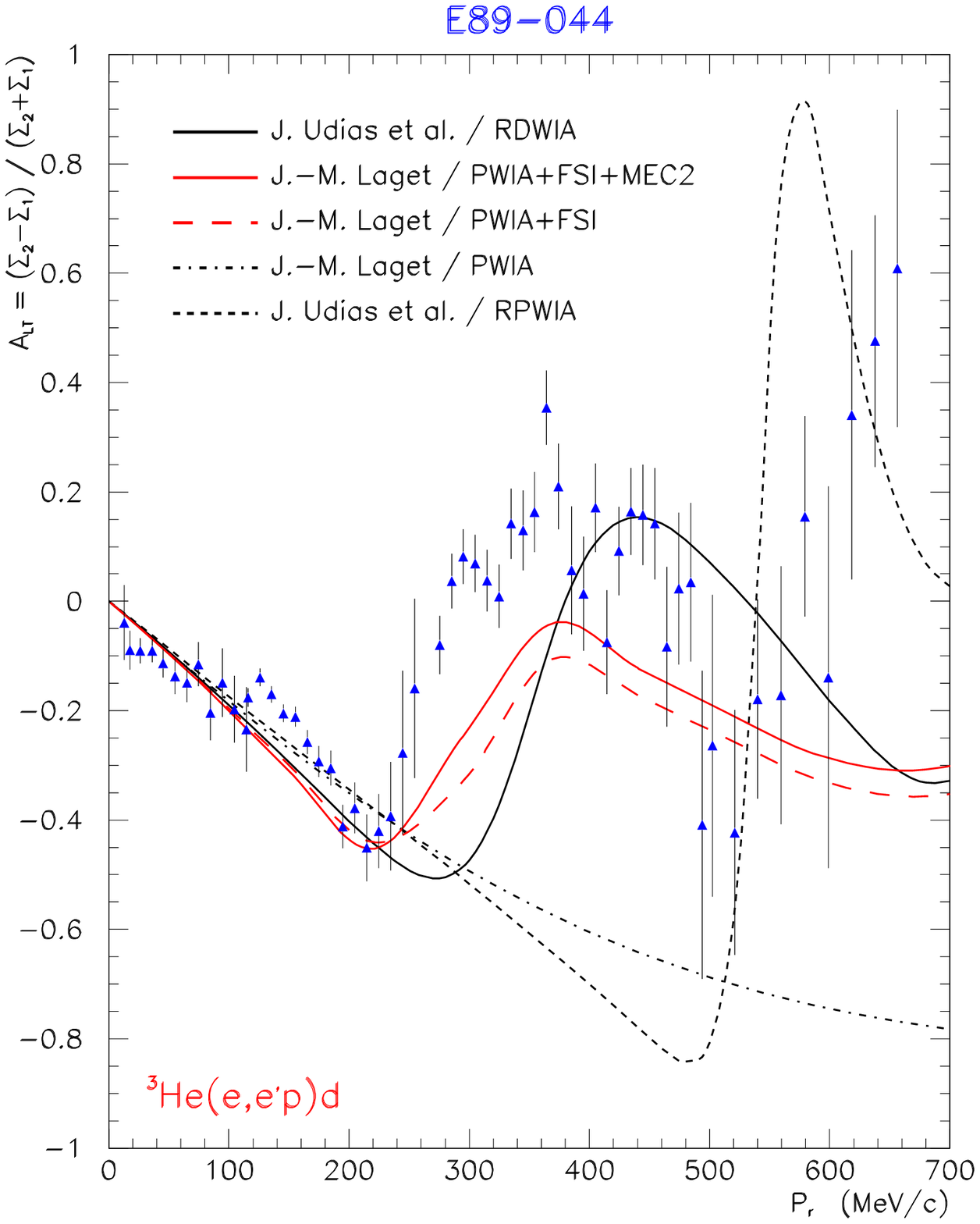, width=185pt} \hspace*{3pt}}
\caption{Left-right asymmetry measurements in the $^{16}$O~\cite{Gao00} (left 
panel) and $^3$He~\cite{Rva04} (right panel) nuclei: the curves show different 
calculations as explained in the text}
\label{alt}
\end{figure}

Figure~\ref{alt} shows the experimental data for this observable measured in 
the E89-003~\cite{E89003} and E89-044~\cite{E89044} experiments for the 
$^{16}$O and $^3$He nuclei, respectively. $A_{LT}$ shows a typical oscillating
pattern characteristic from factorization breakdown which originates however 
from quite different mechanisms in each case. In the $^{16}$O case, different
calculations~\cite{{Kel99},{Udi99},{Cab04}} show that the oscillation comes 
from relativistic corrections to the proton wave function via the enhancement 
of lower spinor components~\cite{Udi01}. In the $^3$He case, FSI are responsible 
for this behaviour~\cite{{Lag04},{Udi04}} which occurs at a similar recoil 
momentum. \newline
It is instructive to compare on the right panel of fig.~\ref{alt} the 
microscopic (J.-M.~Laget) and mean field approach (J.~Udias {\it et al;}) 
calculations. An important feature of the relativistic 
calculation is the breakdown of the cross section factorization at the PWIA 
level~\cite{Udi01} which generates the large oscillation seen at high recoil 
momentum. This feature is not reproduced by the microscopic approach which has 
a smooth variation in PWIA. When taking into account FSI within two different 
methods 
$-$ optical potential for the many-body approach and high energy parametrization 
of the NN scattering for the microscopic one $-$ both calculations get similar 
agreement with data. This not only confirms the importance of FSI but also 
the relative insensitivity of $A_{LT}$ to the nuclear wave function, and would 
suggest that this observable is more sensitive to the elementary current
operator. There are no microscopic calculations for $^{16}$O but the difference 
between distorted non-relativistic and relativistic calculations clearly 
signifies the importance of relativistic corrections to the wave function. It 
is still a mystery how the same observable can be sensitive or insensitive to 
the wave function depending on the nucleus.

The decrease of the nuclear density when going from $^{16}$O to $^3$He has been 
proposed to explain the supression of relativistic effects~\cite{Udi04}. The 
E01-108 experiment~\cite{E01108} will study the $(e,e'p)$ reaction in $^4$He 
for kinematical conditions similar to the $^3$He experiment~\cite{E89044} 
investigating, among other observables, $A_{LT}$. The higher density of the 
$^4$He nucleus will allow this feature to be tested and make a bridge between
many-body and few-body approaches. 

%
%
\section{Nucleon-Nucleon Correlations}
\label{sec:NNcor}

The search for NN correlations with the $(e,e'N)$ and $(e,e'NN)$ reactions
is also an important topic of the Jlab experimental program. At high recoil
momentum, one expects to learn about the short-range repulsive part of the NN 
interaction, and at some point, to abandon the traditional mesonic description 
of the nuclear field in favor of a smaller scale description involving the 
quark substructure of the nucleon.

\subsection{$(e,e'p)$ channel}

One possible experimental method to study this issue is to select a correlated 
pair in the nucleus and measure the recoil momentum distribution~\cite{Mar88}. 
In PWIA, the residual system does not participate in the reaction and
consequently, the $p_r$ distribution directly reflects the initial momentum 
distribution. Within this context the selection of a correlated pair in a 
A$(e,e'p)$ experiment is insured by a strict kinematical relationship bewteen 
$E_m$ and $p_r$
\begin{equation}
E_m = \sqrt{ \left( M_{A-2} + \sqrt{M_N^2+p_r^2} \right)^2 - p_r^2} \, + \, M_p 
\, - \, M_A 
\label{rel:3bbu}
\end{equation}
where the $M_i$ are the nucleons and nuclei mass involved in the process. Thus, 
in PWIA, a signature of the disintegration of a correlated pair is the
appearance of a peak in the cross section as a function of $E_m$, in the 
continuum region, which position depends on $p_r$. 

This experimental method was applied in the E89-044 experiment~\cite{E89044} at 
JLab which has investigated the $^3$He$(e,e'p)np$ 
electro-disintegration of the $^3$He nucleus in the continuum region. Data
taking was performed in perpendicular kinematics up to 1~GeV/$c$, at constant 
electron momentum and angle allowing for a better control of the hadronic 
current. In addition, as compared to previous 
experiments~\cite{{Flo99},{Mar88}}, the $E_m$ acceptance of the JLab experiment 
is large enough to investigate the pair momentum distribution without 
suffering from truncation and subsequent extrapolation of the spectra of the 
relative energy of the pair~\cite{BenTh}. 

\begin{figure}[ht]
\begin{center}
\epsfig{file=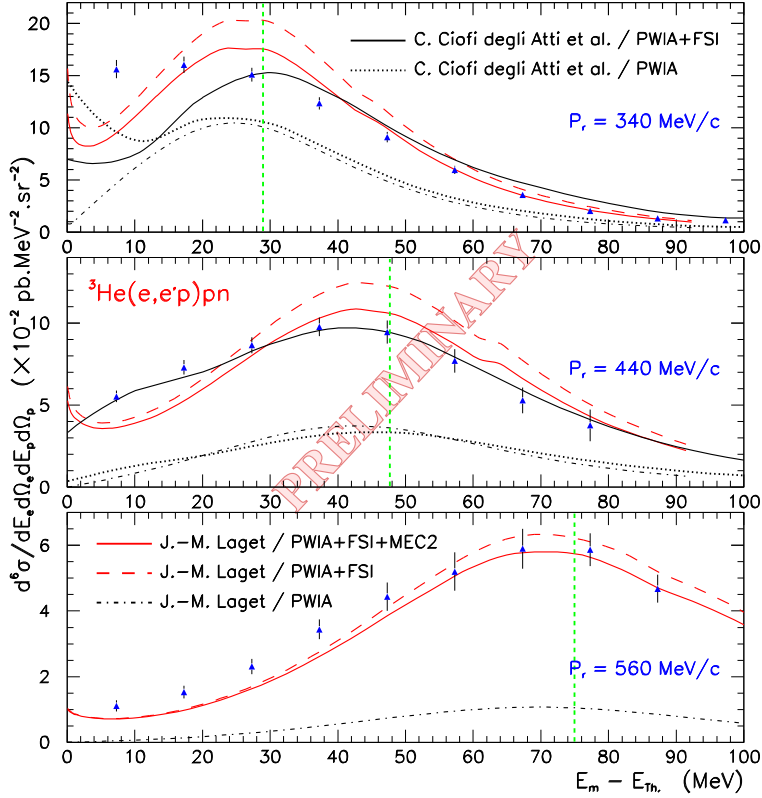, width=320pt}
\caption{Selected data of the E89-044 experiment in the continuum
region~\cite{Ben04}: the green vertical line indiquates the expected location 
of the correlated pair peak; calculations are from C.~Cioffi degli Atti~\cite{Cio04} 
and J.-M.~Laget~\cite{Lag04} for different model approximations.}
\label{3bbu}
\end{center}
\end{figure}
The experimental missing energy distribution for selected recoil momenta is
compared in fig.~\ref{3bbu} to different calculations. A broad peak is 
observed at about the expected location, supporting the observation of a 
correlated pair in $^3$He. However, calculations clearly show that the measured 
strength can only be explained by large FSI contributions and a significant MEC 
correction at small recoil momentum. It originates mainly from the 
re-interaction in the pair and to a lesser extent from the re-interaction with 
the residual nucleon. Therefore, as in the two-body disintegration case 
(sec.~\ref{sec:2bbu}), it is 
not possible to extract an exact value of the actual pair momentum distribution 
because of the small contribution of the PWIA amplitude to the 
reaction cross section. In fact, the relationship of eq.~\ref{rel:3bbu} is 
valid as long as the reaction involves two active nucleons with a spectator 
residual system. When the nucleon of the active pair rescatters, the position,
the width and the shape of the peak might all be affected, the FSI dominance in 
the pair being consistent with previous observations (sec.~\ref{sec:deuton}).

\begin{figure}[ht]
\begin{center}
\epsfig{file=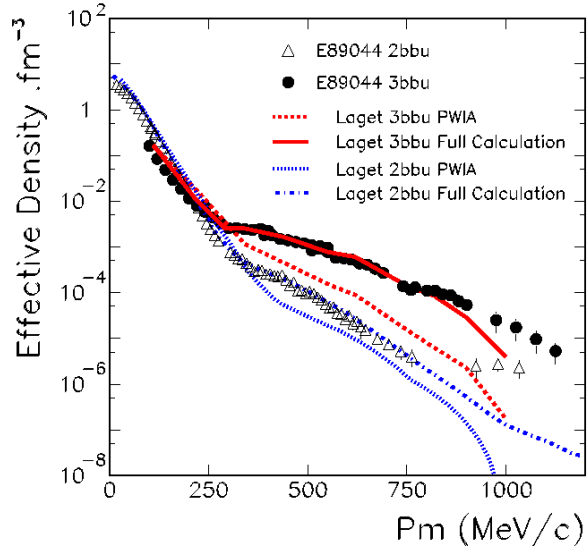, width=240pt}
\caption{Effective nucleon momentum distributions in $^3$He: experimental 
results for the two-~\cite{Rva03} and three-body~\cite{BenTh} breakup are 
compared to calculations from J.M.-Laget~\cite{Lag04}.}
\label{effdens}
\end{center}
\end{figure}
In the abscence of factorization of the cross section, one can 
reconstruct the effective momentum distribution (also known as the reduced 
cross section in the literature), extracted from data by removing the electron 
dependent part. In the particular case of the E89-044 experiment, the two-body 
and three-body breakup channels were simultaneously measured, at the same
electron kinematics. Therefore, the comparison between the two effective
momentum distributions tells about the nuclear structure. Fig.~\ref{effdens} 
particularly suggests that high momentum configurations preferably originate 
from correlated pairs in the nucleus. Using modern potentials and nuclear 
wave functions, theoretical calculations are in good agreement with data, 
and confirm the important role of short-range correlations. 
  
\subsection{$(e,e'pN)$ channel}

While $(e,e'p)$ reactions lead as today to indirect evidence of short-range 
correlations, there are expectations that two-body knockout reactions would provide 
direct and sensitive measurements. To date, the few achieved experiments in many- 
and few-body systems~\cite{{Ond97},{Ond98},{Gro99}} still show strong sensitivity 
to competing reaction mechanisms involving FSI, MEC and $\Delta$ excitation 
(see~\cite{Gra04} for more details).

The E89-027 experiment~\cite{E89027} in hall B of JLab takes advantage of 
the large phase space coverage of CLAS (CEBAF Large Acceptance 
Spectrometer)~\cite{Mec03} to study the $^3$He$(e,e'pN)$ reaction over a wide 
range of energy and momentum transfers. By selecting all nucleon momenta above 
the Fermi momentum and restricting the perpendicular component (transverse to 
$\vec q$) of the fast nucleon to below 300~MeV/$c$, the emphasis is put on 
configurations where the virtual photon interacts with the leading nucleon 
while the associated correlated pair stays at rest~\cite{Niy04}. For such 
configurations, different calculations show that FSI and two-body currents of 
the leading nucleon are suppressed but the continuum state interactions of the 
spectator pair and three-body exchange currents play an important role,
distorting the elementary nuclear information. 

In the near future, with the help of the newly built Big Bite 
spectrometer~\cite{Mon03}, NN correlations will be investigated in 
carbon~\cite{E01105} for anti-parallel kinematics where the PWIA amplitude is
expected to be dominant.

%
%
\section{Bound Nucleon Form Factors}

There are many reasons to at least question the electromagnetic properties of 
bound nucleons as compared to free ones: within the nuclear field, the nucleon 
mass 
changes from its free value to an effective value depending on the intensity of 
the binding force; the bound Dirac spinors describing nucleons differ from free 
ones, the enhancement of the lower spinor components being the basic effect of 
relativistic corrections in $A_{LT}$~\cite{Udi01}; there are three different 
prescriptions for the current operator which are equivalent for on-shell 
nucleons only, and which would strictly lead to twelve different form 
factors for bound nucleons instead of the two usual Dirac and Pauli form 
factors~\cite{Kel96}... Unfortunately, the determination of the modifications 
of nucleon properties can only be done within a model: nucleon changes like the 
polarization of the surrounding meson cloud or the excitation of the quark hard 
core are intertwined with conventional MEC and isobar currents. \newline
An experimental method to minimize these effects and get closer to the 
in-medium nucleon information is the measurement of polarization 
transfer observables~\cite{{Akh74},{Arn81}} in quasi-elastic scattering of
longitudinally polarized electrons. In the free proton case, one can write
\begin{equation}
\frac{G_E}{G_M} \, = \, - \,\frac{P_t}{P_l} \, \frac{E+E'}{2 M_p} \, 
{\tan{ \left( \frac{\theta}{2} \right) }}
\end{equation}
\vspace*{-5pt}
where $P_t$ and $P_l$ are the measured transverse and longitudinal components of 
the proton polarization. This method was successfully applied in the measurement 
of the electric form factor of the free proton~\cite{{Jon00},{Gay02}}, one of 
the most important experimental result of the decade which points to the 
existence of quark orbital momentum. Thanks to these data, the possible 
existence of two photon exchange contributions~\cite{Gui03} strongly questions 
cross section measurements to access the nucleon form factors at high $Q^2$. 
In addition, the polarization transfer method is supposed to be intrinsically 
less sensitive to many-body effects.

\vspace*{10pt}
\begin{figure}[ht]

\vspace*{5pt}

\leftline{\hspace*{5pt}\epsfig{file=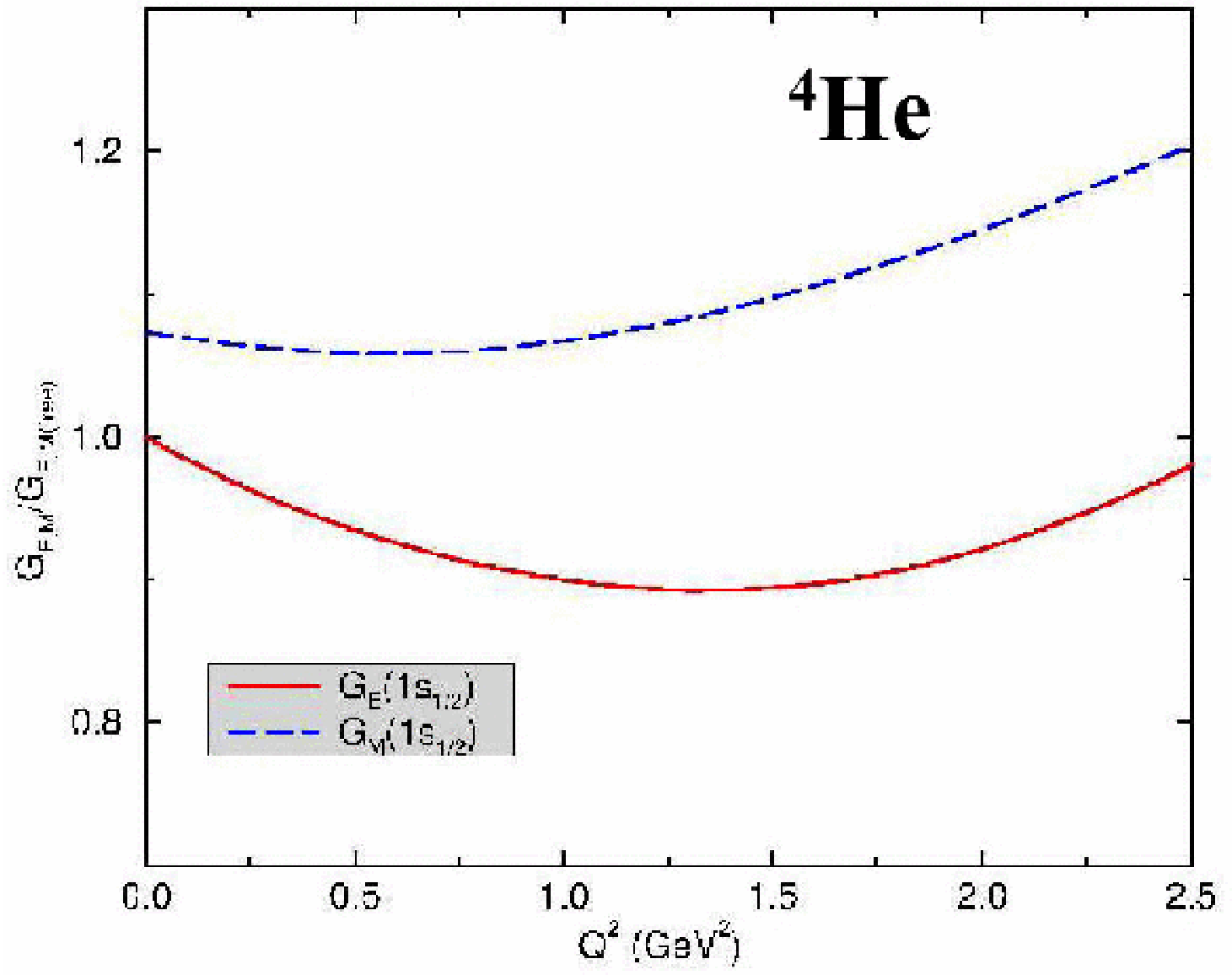, width=160pt}}

\vspace*{-150pt}
 
\rightline{\epsfig{file=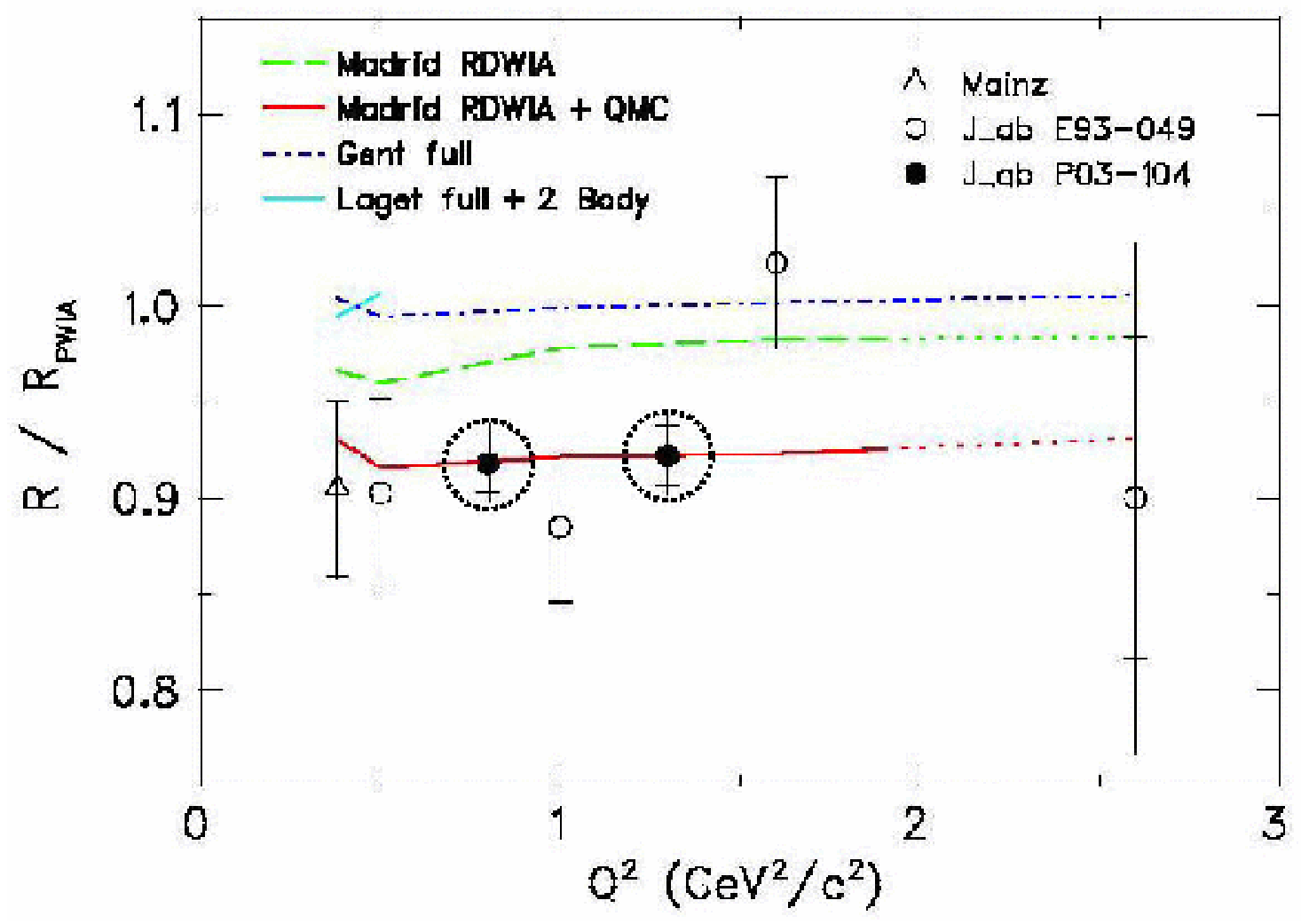, width=220pt}\hspace*{5pt}}

\caption{The modification of in-medium proton form factors: theoretical
expectations in $^4$He~\cite{Lu99} (left panel), and measured~\cite{Str03} and 
projected~\cite{E03104} experimental results (right panel). The latter are 
expressed in terms of the ratio of the measured polarization ratio 
$R=(P_t/P_l)_{^4{\mathrm He}} \big/ (P_t/P_l)_{\mathrm H}$ normalized by the free 
value, to the same ratio calculated in PWIA.}
\label{bff}
\end{figure}
Following a previous measurement at small $Q^2$ in MAMI~\cite{Die01}, the JLab 
E93-049 experiment~\cite{E93049} had investigated this problem in the 
$^4$He$({\vec e},e'{\vec p})^3$H quasi-elastic scattering where theoretical
calculations predict a sizeable change of the proton electromagnetic form 
factors, originating mainly from the enhancement of the lower 
spinor components of the in-medium quark wave function~\cite{Lu99}. Experimental 
data~\cite{Str03} are shown in fig.~\ref{bff} together with several 
calculations. The most modern calculations considering 
free form factors and many-body dynamics fail to reproduce data. Agreement is 
only obtained after including in these calculations the predicted 
modifications~\cite{Lu99}. The authors~\cite{Str03} claim for an evidence of 
medium modifications. However, the accuracy of the data does not yet allow a 
definitive conclusion. It is the purpose of the E03-104 experiment~\cite{E03104} 
to provide high accuracy data (fig.~\ref{bff}) in this momentum range and 
hopefully confirm this finding.

%
%
\section{Conclusions}

To date, the $(e,e'p)$ JLab experimental program on few-body systems has 
established the dominance of FSI on top of the quasi-elastic peak. FSI are 
responsible for moderate quenching ($p_r < 300$~MeV/$c$) and large enhancement 
($p_r > 300$~MeV/$c$) of the cross section, and result in a large oscillation 
in $A_{LT}$. The same mechanism also obscures the signal from correlated 
nucleon pairs. This all confirms that the high momentum components of the 
nuclear wave function cannot be measured in such kinematics, but hopefully 
more reliably at large $x$, as predicted by models. \newline
The dominance of FSI at $x=1$ opens a new field of investigation. This feature 
can be advantageously used to study CT in few-body systems which should 
manifest as a function of $Q^2$ by a reduction of the cross section at large 
recoil momentum ($p_r > 400$~MeV/$c$). \newline
Finally, there are indications that the electromagnetic properties of nucleons 
are modified in the nuclear medium. The still on-going activity in this domain,
as well on the analysis front as on the future experimental program, will 
confirm the status of these modifications. 
 
%
%
\section*{Acknowledgements}

I would like to thank the organizers of the XXIIrd International Workshop on
Nuclear Theory for their invitation and warm hospitality at Rila Mountain. 

This work was supported in part by the U.S. Department of Energy (DOE) 
contract DE-AC05-84ER40150 Modification No. M175 under which the Southern 
Universities Research Association (SURA) operates the Thomas Jefferson National 
Accelerator Facility, the National Science Foundation, the Italian Istituto 
Nazionale di Fisica Nucleare (INFN), the French Atomic Energy Commission and 
National Center of Scientific Research, and the Natural Science and Engineering 
Research Council of Canada. 

%
%

%
%

\begin{thebibliography}{99}
%
\bibitem{Fru84} 
S.~Frullani and J.~Mougey, {\it Adv. Nucl. Phys.} {\bf 14} (1984) 1.
%
\bibitem{Kel96} 
J.J.~Kelly, {\it Adv. Nucl. Phys.} {\bf 23} (1996) 75.
%
\bibitem{Ulm02}
P.~Ulmer {\it et al.}, {\it Phys. Rev. Lett.} {\bf 89} (2002) 062301.
%
\bibitem{Rva04}
M.M.~Rvachev, F.~Benmokhtar, E.~Penel-Nottaris {\it et al.}, {\it 
nucl-ex/0409005}, (2004). 
%
\bibitem{Rei03}
B.~Reitz, Proc of the {\it International Workshop on Probing Nucleons and 
Nuclei with the $(e,e'p)$ reaction}, Edts. E.~Voutier, J.-M.~Laget, 
D.~Higinbotham, Grenoble (France), 14-17 October 2003.
%
\bibitem{JLa12}
A.~Afanasev {et al.}, {\it Hall A - 12~GeV upgrade: pre-conceptual design 
report}, Jefferson Laboratory, (2002).
%
\bibitem{Gao00}
J.~Gao {et al.}, {\it Phys. Rev. Lett.} {\bf 84} (2000) 3265.
%
\bibitem{Niy04}
R.A.~Niyazov, L.B.~Weinstein {\it et al.}, {\it Phys. Rev. Lett.} {\bf 92} 
(2004) 052303.
%
\bibitem{Ben04}
F.~Benmokhtar, M.M.~Rvachev, E.~Penel-Nottaris {\it et al.}, {\it 
nucl-ex/0408015}, (2004).
%
\bibitem{Str03}
S.~Strauch, S.~Dieterich {\it et al.}, {\it Phys. Rev. Lett.} {\bf 91} (2003) 
052301.
%
\bibitem{For83}
T.~de~Forest, {\it Nucl. Phys.} {\bf A 392} (1983) 232.
%
\bibitem{Alc04} 
J.~Alcorn, {\it Nucl. Inst. Meth. Phys. Res.} {\bf A 522} (2004) 294.
%
\bibitem{Jon94}
Jefferson Lab Experiment {\bf E94-004}, M.~Jones, P.~Ulmer, spokespeople, 
(1994).
%
\bibitem{Lag04}
J.-M.~Laget, {\it nucl-th/0407072}, (2004).
%
\bibitem{E89044}
Jefferson Lab Experiment {\bf E89-044}, M.~Epstein, A.~Saha, E.~Voutier, 
spokespeople, (1989).
%
\bibitem{Jan82}
E.~Jans {\it et al.}, {\it Phys. Rev. Lett.} {\bf 49} (1982) 974. 
%
\bibitem{Flo99}
R.~Florizone {\it et al.}, {\it Phys. Rev. Lett.} {\bf 83} (1999) 2308. 
%
\bibitem{Kie97}
A.~Kievsky, E.~Pace, G.~Salme, M.~Viviani, {\it Phys. Rev.} {\bf C 56} (1997) 64. 
%
\bibitem{Lag03}
J.-M.~Laget, {\it Few Body Syst. Supp.} {\bf 15} (2003) 171.
%
\bibitem{Cio04}
C.~Cioffi~degli~Atti and L.~Kaptari, {\it nucl-th/0407024}, (2004).
%
\bibitem{Udi04}
J.~Udias and J.R.~Vignote, Cont. to the {\it XXIIIrd International Workshop on
Nuclear Theory}, Rila Mountain (Bulgaria) 14-19 June, 2004.
%
\bibitem{Maz03}
M.~Mazouz, Diploma Dissertation, Ecole Nationale Sup\'erieure de Physique,
Grenoble (France), 2003.
%
\bibitem{Rva03}
M.~Rvachev, Ph.D. Thesis, Massachusetts Institute of Technology, Cambridge (MS, 
USA), 2003.
%
\bibitem{E97111}
Jefferson Lab Experiment {\bf E97-111}, J.~Mitchell, B.~Reitz, J.~Templon,
spokespeople, (1997).
%
\bibitem{Tad87}
S.~Tadokoro, T.~Katayama, Y.~Akaishi, and H.~Tanaka, {\it Prog. Theor. Phys.} 
{\bf 78} (1987) 732.
%
\bibitem{Lee98}
J.J~van~Leeuwe {\it et al.}, {\it Phys. Rev. Lett.} {\bf 80} (1998) 2543. 
%
\bibitem{Pen04}
E.~Penel-Nottaris, Doctorat Thesis, Universit\'e Joseph Fourier, Grenoble
(France), 2004.
%
\bibitem{E01020}
Jefferson Lab Experiment {\bf E01-020}, W.~Boeglin, M.~Jones, A.~Klein, 
J.~Mitchell, P.~Ulmer, E.~Voutier, spokespeople, (2001).
%
\bibitem{Boe03}
W.~Boeglin, Proc of the {\it International Workshop on Probing Nucleons and Nuclei 
with the $(e,e'p)$ reaction}, Edts. E.~Voutier, J.-M.~Laget, D.~Higinbotham, 
Grenoble (France), 14-17 October 2003.
%
\bibitem{Bia96}
A.~Bianconi, S.~Jeschonnek, N.N.~Nikolaev and B.G.~Zakharov, {\it Phys. Rev.} 
{\bf C 53} (1996) 576.
%
\bibitem{Fra97}
L.L.~Frankfurt, M.M.~Sargsian and M.L.~Strikman, {\it Phys. Rev.} {\bf C 56}
(1997) 1124.
%
\bibitem{Vou96}
E.~Voutier {\it et al.}, Proc. of the {\it Second ELFE Workshop 
on Hadronic Physics}, Saint-Malo (France), Edts. N.~d'Hose, B.~Frois, 
P.A.M.~Guichon, B.~Pire and J.~van~de~Wiele, DAPNIA-SPhN-{\bf 96-35} (1996) 
107.
%
\bibitem{Lag97}
J.-M.~Laget, Proc. of the {\it CT97, Workshop on Color Transparency}, Edt. 
E.~Voutier, Grenoble (France), 25-27 June 1997, (1998) 131.
%
\bibitem{One95} 
T.G.~O'Neill {\it et al.}, {\it Phys. Lett.} {\bf 87} (1995) 87.
%
\bibitem{Abb98} 
D.~Abbot {\it et al.}, {\it Phys. Rev. Lett.} {\bf 80} (1998) 5072.
%
\bibitem{Gar02} 
K.~Garrow, D.~McKee {\it et al.}, {\it Phys. Rev.} {\bf C 66} (2002) 044613.
%
\bibitem{Gil98} 
S.~Gilad, W.~Bertozzi, and  Z.L.~Zhou, {\it Nucl. Phys.} {\bf A 631} (1998) 
276c.
%
\bibitem{Udi01}
J.~Udias, J.~Javier, E.M.~de~Guerra, A.~Escuderos and J.~Caballero, Proc. of the 
{\it Vth Workshop on Electromagnetic Induced Two-Hadron Emission}, Lund (Sweden], 
2001; {\it nucl-th/0109077}, (2001). 
%
\bibitem{E89003}
Jefferson Lab Experiment {\bf E89-003}, W.~Bertozzi, K.~Fissum, A.~Saha, 
L.~Weinstein, spokespeople, (1989).
%
\bibitem{Kel99}
J.J.~Kelly, {\it Phys. Rev.} {\bf C 60} (1999) 044609.
%
\bibitem{Udi99}
J.M.~Udias, J.A.~Caballero, E.~Moya~de~Guerra, J.E.~Amaro and T.W.~Donnelly, 
{\it Phys. Rev. Lett.} {\bf 84} (1999) 5451.
%
\bibitem{Cab04}
J.A.~Caballero, Cont. to the {\it XXIIIrd International Workshop on Nuclear 
Theory}, Rila Mountain (Bulgaria) 14-19 June, 2004.
%
\bibitem{E01108}
Jefferson Lab Experiment {\bf E01-108}, K.~Aniol, S.~Gilad, D.~Higinbotham, 
A.~Saha, spokespeople, (2001).
%
\bibitem{Mar88}
C.~Marchand {\it et al.}, {\it Phys. Rev. Lett.} {\bf 60} (1988) 1703.
%
\bibitem{BenTh}
F.~Benmokhtar, Ph.D. Thesis, Rutgers, The State University of New Jersey, 
Piscataway (NJ, USA), 2004.
%
\bibitem{Ond97}
G.~Onderwater {\it et al.}, {\it Phys. Rev. Lett.} {\bf 78} (1997) 4893.
%
\bibitem{Ond98}
G.~Onderwater {\it et al.}, {\it Phys. Rev. Lett.} {\bf 81} (1998) 2213.
%
\bibitem{Gro99}
D.L.~Groep {\it et al.}, {\it Phys. Rev. Lett.} {\bf 83} (1999) 5443.
%
\bibitem{Gra04}
P.~Grabmayr, Cont. to the {\it XXIIIrd International Workshop on
Nuclear Theory}, Rila Mountain (Bulgaria) 14-19 June, 2004.
%
\bibitem{E89027}
Jefferson Lab Experiment {\bf E89-027}, W.~Bertozzi, W.~Boeglin, L.~Weinstein, 
spokespeople, (1989).
%
\bibitem{Mec03} 
B.~Mecking, {\it Nucl. Inst. Meth. Phys. Res.} {\bf A 503} (2003) 513.
%
\bibitem{Mon03}
P.~Monaghan, Proc of the {\it International Workshop on Probing Nucleons and 
Nuclei with the $(e,e'p)$ reaction}, Edts. E.~Voutier, J.-M.~Laget, 
D.~Higinbotham, Grenoble (France), 14-17 October 2003.
%
\bibitem{E01105}
Jefferson Lab Experiment {\bf E01-105}, W.~Bertozzi, E.~Piasetzky, 
J.W.~Watson, S.~Wood, spokespeople, (2001).
%
\bibitem{Akh74} 
A.I.~Akhiezer, M.P.~Rekalo {\it Sov. Jour. Part. Nucl.} {\bf 3} (1974) 274.
%
\bibitem{Arn81} 
R..~Arnold, C.~Carlson, F.~Gross, {\it Phys. Rev.} {\bf C 23} (1981) 363.
%
\bibitem{Jon00}
M.~Jones {\it et al.}, {\it Phys. Rev. Lett.} {\bf 84} (2000) 1398.
%
\bibitem{Gay02}
O.~Gayou {\it et al.}, {\it Phys. Rev. Lett.} {\bf 88} (2002) 092301.
%
\bibitem{Gui03}
P.A.M.~Guichon and M.~Vanderhaeghen, {\it Phys. Rev. Lett.} {\bf 91} (2003) 
142303.
%
\bibitem{Die01}
S.~Dieterich {\it et al.}, {\it Phys. Lett.} {\bf B 500} (2001) 47.
%
\bibitem{E93049}
Jefferson Lab Experiment {\bf E93-049}, R.~Ent, P.~Ulmer, spokespeople, (1993).
%
\bibitem{Lu99}
D.H.~Lu, K.~Tsushima, A.W.~Thomas, A.G.~Williams and K.~Saito, {\it Phys. Rev.} 
{\bf C 60} (1999) 068201.
%
\bibitem{E03104}
Jefferson Lab Experiment {\bf E03-104}, R.~Ent, R.~Ransome, S.~Strauch, 
P.~Ulmer, spokespeople, (2003).
%
\end{thebibliography}
\end{document}